\documentstyle[11pt]{article}
 
\def\pa{\partial}
\def\k{\kappa} 
\def\g{\gamma} \def\G{\Gamma}
\def\a{\alpha} 
\def\b{\beta} 
\def\d{\delta} \def\D{\Delta}
\def\e{\epsilon} 
 
\def\k{\kappa}
\def\l{\lambda} 
\def\m{\mu}  
\def\n{\nu}

\def\r{\rho} 
\def\s{\sigma} 

\def\t{\tau}

\def\mn{{\mu \nu}}
\def\ab{{\alpha \beta}}
\def\be{\begin{equation}}
\def\ee{\end{equation}}
     
\begin{document}

\begin{flushright} 
BRX TH--450\\
LPTENS--99/14
\end{flushright}

\begin{center}
{\Large\bf Graviton-Graviton Scattering,\\
Bel--Robinson and Energy (Pseudo)-Tensors}
\end{center}

\renewcommand{\thefootnote}{\fnsymbol{footnote}}
\setcounter{footnote}{0}
\begin{center}
S. Deser$^{(a)}$\footnote{\tt deser@brandeis.edu,
jfrankli@aprilia.stanford.edu,
domenico.seminara@lpt.ens.fr},
J.S. Franklin$^{(a)*}$
\footnote{Present
address: Stanford University,
PO Box 13236, Stanford, CA 94309, USA}
and
D. Seminara$^{(a,b)*}$

\vspace{.1in}

$^{(a)}$ Department of Physics, Brandeis University \\
Waltham, MA 02454, USA\\
$^{(b)}$ Laboratoire de Physique Th{\'e}orique, 
{\'E}cole Normale Sup{\'e}rieure.\\
24 rue Lhomond, F-75231, Paris CEDEX 05, France.
\end{center}

\begin{quotation}
Motivated by recent work involving the graviton-graviton
tree scattering amplitude, and its twin descriptions
as the square of the Bel--Robinson tensor, 
$B_{\m\n\a\b}$, and as the ``current-current
interaction" square of gravitational
energy pseudo-tensors $t_{\a\b}$,
we find an exact tensor-square root equality 
$B_{\mn\a\b} = \pa^2_\mn t_{\a\b}$, for a combination
of Einstein and Landau--Lifschitz $t_\ab$, in Riemann
normal coordinates.  In the process, we relate, on-shell,
the usual superpotential basis for classifying
pseudo-tensors with one spanned by polynomials in the
curvature.
\end{quotation}

\renewcommand{\thefootnote}{\arabic{footnote}}
\setcounter{footnote}{1}

\renewcommand{\theequation}{\arabic{equation}}
\setcounter{equation}{1}

\noindent{\bf 1. Introduction}

This paper revisits and relates the three ancient
subjects of our title in a novel way, suggested to
us in a very different context, the construction of
invariants of D=11 supergravity \cite{001}.  
The graviton-graviton tree level
amplitude $M$, particularly in the present 
exclusively D=4 context, is remarkably simple, namely
the square of the
famous Bel--Robinson tensor \cite{002a},
$$
L  \sim B^2_{\mn\a\b}\; . \eqno{(1{\rm a})}
$$
To be sure, there is some work involved here:
First, since $M$ is generated by
intermediate graviton exchange, it is nonlocal, but
when multiplied by the Mandelstam variables
$stu$, it reduces to a local invariant $L$.  The
latter must, by power counting and (abelian) gauge
invariance to this lowest 
$(\k h_\mn )^4 \equiv (g_\mn - \eta_\mn )^4$ order, be
a scalar proportional to $R^4_{\mn\a\b}$.  [The 
external gravitons are all on-shell $(R_\mn = 0)$, so
$R_{\mn\a\b}$ really means the Weyl tensor throughout.]
Here the power of using a suitable
basis to classify scalar or tensorial powers of
curvature \cite{002} first emerges.  For example, 
there are seven
algebraically independent $R^4$ scalar monomials in 
generic dimension, but these can be shown to reduce
to just two in D=4, owing to identities such as
$R_{\m\a\b\g} R^{\n\a\b\g} \equiv 
\textstyle{\frac{1}{4}} \d^\n_\m R^2_{\a\b\g\d}$
valid only there.  This basis is spanned by
$(E^2, P^2)$, where $(E,P)$ are the D=4 Euler and Pontryagin
invariants.  Since graviton scattering must be maximally
helicity conserving (as follows directly from the 
supersymmetrizability of Einstein theory), this singles
out $L \sim (E-P)(E+P)$; but another deep D=4
fact is that this is in turn just $B^2$.

The final objects in our title, 
the gravitational energy
pseudo-tensors, enter in a very physical way as the
currents that generate the scattering
through the cubic vertex
$\sim \k \int d^4x h \pa h \pa h$ in the expansion of the
Einstein action, which can be expressed as the coupling of
the field $h_{\a\b}$ to a current $t^{\a\b} \sim \k
(\pa h \pa h)^{\a\b}$.  The amplitude
can then be expressed in the usual current-current form
$M \sim \int t^{\l\s} D_{\l\s\a\b} t^{\a\b}$, where $D$
is the intermediate graviton's propagator. Although 
gauge invariance
further requires inclusion of the quartic contact term
$\k^2 \int d^4x hh \pa h\pa h$,  it is 
plausible that this characterization of $M$ yields
another version of $L$,
$$
L \sim ( \pa^2_\mn t_{\a\b})^2 \eqno{(1{\rm b})}
$$
upon keeping track of momentum dimensions.  
Hence the temptation
to equate the ``tensor square roots" of (1a) and (1b),
\be
B_{\mn\a\b} \stackrel{?}{=} \pa^2_\mn \, t_{\a\b} \; .
\ee

Let us next recall some properties of $t_{\a\b}$ and
earlier attempts to relate $B$ and $t$.  First, we
note that gauge systems with spin $>$1 do not even possess
gauge invariant stress tensors, but only integrated
Poincar\'{e} generators \cite{DandM}.  While this local defect
is irrelevant in the absence of gravity, it becomes
critical when the stress tensor is to be its source,
not least for (self-coupled) general relativity itself, as
witness the history of 
its countless energy pseudotensors (see {\it e.g.},
\cite{005}).
The problem is of course the clash between the second 
derivative order required to build the (linearized) in-
or (generically) co-variant curvatures and the first 
derivative order building blocks of the $t_{\a\b}$.
Indeed, the Bel--Robinson tensor arose as 
an attempt at constructing
a ``more covariant" tensorial quantity modelled on the Maxwell 
stress tensor. However, the price is high:
$B$ has too many
derivatives and indices, it is only covariantly conserved
and it is not a physical current \cite{006}.  Clearly any putative
relation like (2), even at linearized level, must be a
non-invariant one.  Nevertheless, there exists a textbook
exercise \cite{002b} proposing the equality 
$B_{\mn\a\b} = \pa^2_\mn t_{\a\b} + S_{\mn\a\b}$
in Riemann normal coordinates (RNC) for a particular (Einstein)
$t_{\a\b}$, but with a rather mysterious remainder.
In achieving (2), we will have to consider all possible
pseudo-tensors; as we have mentioned, there is an infinity of 
them and being identically ordinarily 
(rather than covariantly) conserved on shell, they 
can all be expressed as superpotentials there.  This is also the
case for $B$: on shell, it reduces to the identically
(covariantly) conserved Sachs tensor \cite{004}.
In establishing (2), in the particular RNC frame,
we will in fact translate the superpotential ``basis"
for the $t_{\a\b}$ and for $B$ into one for 
four-index curvature quadratics.  This will enable
us not only to verify that it can be done there, but also
how uniquely.

Despite our arguments in favor of the existence of an exact 
connection (2) for some $t_{\a\b}$, its validity is
far from obvious; $B$ is totally symmetric and traceless,
while the $\pa^2_\mn t_{\a\b}$ do not  even
display
$(\a\b )$ symmetry in general, let alone the other
invariances of $B$.  It is therefore  amusing 
that this correspondence, even if highly gauge
variant, can be established at all!

\noindent{\bf 2. Ingredients}

\noindent A. Bel--Robinson. ~As mentioned, $B$
first appeared in the endless search for a  
covariant version of 
gravitational energy density; the analogy 
with the Maxwell stress tensor  
$T_\mn = F_{\m\a}F_\n~^\a + ^*\!F_{\m\a}~
^*\!F_\n~^\a$  ($^*\!F$ is the dual field strength)
is striking in either of the two equivalent expressions,
\begin{equation}
\label{BR4}
B_{\mu\nu\alpha\beta}=
R^{\rho\ \sigma\ }_{\ \mu\ \alpha} R_{\rho\nu\sigma\beta}+
^*\!R^{\rho\ \sigma\ \ ~*}_{\ \mu\ \alpha} R_{\rho\nu\sigma\beta}\;,
\end{equation}
\begin{equation}
\label{BRD1}
B_{\mu\nu\alpha\beta}=R^{\rho\ \sigma\ }_{\ \mu\ \alpha} 
R_{\rho\nu\sigma\beta}+R^{\rho\ \sigma\ }_{\ \mu\ \beta} 
R_{\rho\nu\sigma\alpha}-\textstyle{\frac{1}{2}} g_{\mu\nu} 
R_{\alpha}^{\ \rho\sigma\tau}R_{\beta\rho\sigma\tau} \; . 
\end{equation}
Here the dual curvature is 
$^*\!R^\mn~_{\l\s} \equiv \textstyle{\frac{1}{2}}
\e^\mn~_{\a\b} R^{\a\b}~_{\l\s}$.
The interplay between these two expressions underlies
the various special properties that $B$ enjoys in
four dimensions, due to on-shell identities that 
(like the dual itself) are only valid
there.  These include, besides the $(\m \leftrightarrow \n )$
and $(\a \leftrightarrow \b )$ symmetries, two further ones:
~(a) $(\mu\nu)\leftrightarrow(\alpha\beta)$ symmetry. 
This is transparent in (\ref{BR4}); 
requiring it in (\ref{BRD1}) demands that
\begin{equation}
g_{\mu\nu} 
R_{\alpha}^{\ \rho\sigma\tau}R_{\beta\rho\sigma\tau}=
g_{\alpha\beta} 
R_{\mu}^{\ \rho\sigma\tau}R_{\nu\rho\sigma\tau} \; ,
\end{equation}
{\it i.e.}, that $R_\a~^{\b\g\t}R_{\b\r\s\t}$ be
a pure trace:
\begin{equation}
\label{rel1}
R_{\alpha}^{\ \rho\sigma\tau}R_{\beta\rho\sigma\tau}-
\textstyle{\frac{1}{4}}g_{\alpha\beta} 
R^{\mu\rho\sigma\tau}R_{\mu\rho\sigma\tau} = 0 \;.
\end{equation} 
Thus, this well-known D=4 identity is encoded in $B$. 
~(b) $(\mu\leftrightarrow\nu)$ symmetry: 
``dualizing" $B$ with 
$\e^{\g\t\mn}$ in (\ref{BR4})
obviously annihilates it by the 
``$^{**} = -1$" property.  In terms of 
(\ref{BRD1}), this then
implies the more general 4-index relation
\begin{equation}
\label{rel2}
\textstyle{\frac{1}{2}}R_{\mu\alpha}\ ^{\rho\sigma}
R_{\nu\beta\rho\sigma}
+R^{\rho\ \sigma\ }_{\ \mu\ \beta} 
R_{\rho\nu\sigma\alpha}-R^{\rho\ \sigma\ }_{\ \nu\ \mu} 
R_{\rho\alpha\sigma\beta}-\textstyle{\frac{1}{8}}
(g_{\mu\nu} g_{\alpha\beta}
- g_{\mu\beta} g_{\nu\alpha})R_{\lambda\rho\sigma\tau}
R^{\lambda\rho\sigma\tau}=0 
\end{equation}
of which (\ref{rel1}) is the $(\a\b )$ trace; indeed these
identities ``explain" why $B$ is necessarily 4-index,
rather than 2-index like $t_{\a\b}$. [For D$>$4 
the identities (6),(7) cease to hold and 
a three-parameter family of 
conserved $B$-like generalization (each
with partial properties)
can be  constructed; for
more about $B$'s and the curvature quartics $BB$ in
D$>$4, see \cite{001}.]

\noindent B. Riemann Normal Coordinates (RNC). ~We recall 
that at any one point in a
Riemann space, coordinate invariance can be exploited 
in order to simultaneously  ~(i) rotate the metric
to Minkowski form
(local flatness), ~(ii) annihilate all affinities (free
fall) and ~(iii) remove 80 of the 100 $\pa^2_\mn
g_{\a\b}$ by use of the 80 independent
$\pa^3_{\a\b\g} \xi_\d$ components of the gauge functions
$\xi_\d$ in the curvature, 
leaving only 10 $\pa^2 g$ combinations to
represent the 10 Weyl curvatures $R_{\mn\a\b}$ of interest
on shell.  
In addition to (i) and (ii),
\be
g_\mn = \eta_\mn \;, \;\;\;\;\;\;
g^\mn = \eta^\mn \; , \;\;\;\;\;\;
g_{\mn , \a} = 0 \; ,
\ee
the choices (iii) defining RNC are summarized as follows: 
\be
-3 \: g_{\mn ,\a\b} = R_{\m\a\n\b} + R_{\m\b\n\a} \; ,
\;\;\;\;\;\;\;\;
-3 \: \G^\a_{\mn ,\b} = R^\a~_{\m\n\b} + R^\a~_{\n\m\b} \; .
\ee
Note that raising and lowering indices 
``passes through derivatives", here denoted by commas. On-shell, 
this means in addition, that all traces of (9) vanish:
\be
g^\mn g_{\mn ,\a\b} = 0 \; , \;\;\;\;\;\;
g^\mn \G^\a_{\mn , \b} = 0 \; , \;\;\;\;\;\;
\G^\a_{\a\b , \m} = 0 = \G^\a_{\mn ,\a} \; .
\ee

\noindent C. Pseudo-tensors. As we have stated, these
can be parametrized either by all possible independent
superpotentials (of second derivative order), or more usefully
in RNC by the appropriate curvature basis.  Before proceeding
to the general case, however, we write directly the two 
specific $t_{\a\b}$ that will enter in (2), namely
the non-symmetric Einstein $E^\a_\b$ and symmetric
Landau--Lifschitz $L^{\a\b}$.  This will also give an
idea of how RNC simplifies matters.  We only keep those
terms in each that will not vanish on-shell because
of  (10)
after taking the two further, $\pa^2_\mn$, derivatives:
$$
E^\a_\b = -2 \G^\a_{\l\s} \: \G^\l_{\b\s} + \d^\a_\b
\G^\l_{\s\t} \: \G^\t_{\s\l} 
\eqno{(11{\rm a})}
$$
$$
L^\ab = - \G^\l_{\a\s}\G^\s_{\b\l} + \G^\a_{\l\s}
\G^\b_{\l\s} -
(\G^\s_{\a\l}\G^\b_{\s\l} + \G^\s_{\b\l} \G^\a_{\s\l} )
+ g^\ab \G^\l_{\s\t} \G^\s_{\l\t} \; . \eqno{(11{\rm b})}
$$
In writing these expressions, we have used the economy
of notation permitted by RNC: first, since both $E$ and
$L$ are special in being bilinear in $\G$'s
({\it i.e.}, not involving terms like $g\pa^2 g$), then both
of the $\pa^2_\mn$ must act on these $\G$'s and not on any
undifferentiated 
metrics; we have therefore set all the latter
to their Minkowski values, so that summation and moving
indices in (11) is to be understood in light of
(8).  Note also that, unlike $L^\ab$, $E^\a_\b$ 
(here expressible as $E^\ab$)
is not yet $(\ab )$-symmetric;
this will, however, come about after differentiation.

\renewcommand{\theequation}{\arabic{equation}}
\setcounter{equation}{11}

\noindent D. Quadratic Curvature Basis.  All 4-index curvature
bilinears are algebraically equivalent to the double contraction
\be
Q_{\m\a\n\b} \equiv R_{a \m b\a} R_{a\n b\b} =
Q_{\n\b\m\a} = Q_{\a\m\b\n} \; ,
\ee
where we have indicated the symmetries of $Q$;
there are no further ones, either within
a pair or under index exchange between pairs.  The three
basis members we need are representable by
$$
X_{\mn\a\b} \equiv  \: Q_{\a\m\n\b} \;\;\;\;\;
Y_{\mn\a\b} \equiv  \: Q_{\a\b\mn} \;\;\;\;\;
Z_{\mn\a\b} \equiv Q_{\a\m\b\n} \;\;\;\;\;\;\;\;(+\n\m )\; , 
\eqno{(13{\rm a})}
$$
where $(+\n\m )$ means that each quantity is to
include the $(\n\m )$-interchanged form in its definition.
These three linearly independent
quantities are a subset of the
complete (6-member) non-symmetrized basis and agree with
the enumerations in \cite{002}.  It is convenient,
in addition, to define a single trace object, which we
take to be
$$
T_{\mn\a\b}\equiv -\textstyle{\frac{1}{6}} \: g_\mn \;
Q_{\s\a \s\b} = - \textstyle{\frac{1}{24}}
\: g_\mn g_{\a\b} \: R^2_{\l\s\g\d} \; .
\eqno{(13{\rm b})}
$$
All the ({\it X,Y,Z,T}) are uniformly labelled by the
indices ($\mn\a\b$) in that order; we will avoid 
index proliferation below by (usually) leaving them off
altogether.
In terms of this basis, $B_{\mn\a\b}$ is the 
combination
\renewcommand{\theequation}{\arabic{equation}}
\setcounter{equation}{13}
\be
B = Z  + 3T  \; .
\ee
It is also convenient to define the tensor 
$S_{\mu\nu\alpha\beta}
\equiv R_{\mu\alpha}^{\ \ \rho\sigma}
R_{\nu\beta\rho\sigma}+R_{\nu\alpha}^{\ \ \rho\sigma}
R_{\mu\beta\rho\sigma}+\frac{1}{4} g_{\mu\nu} 
g_{\alpha\beta}R^2_{abcd}$ which is expressible as
\be
S = 2 B  - 2X -12 T  \; .
\ee
It also follows by (7) that
\be
S = 2Y -2 B -12T + 6 (T_{\m\a\n\b} +
T_{\m\b\n\a} )\; .
\ee

\noindent{\bf 3. The Relation} 

~In RNC, any $\pa^2_\mn
t_{\a\b}$ will be a sum of products 
$\sim (\pa^2 g \: \pa^2 g)_{\a\b\mn}$, with manifest
$(\m\n )$, but not necessarily $(\a\b )$, symmetry.
There is thus a small class of possible terms; all are
expressible in terms of the basis
$(X,Y,Z,T)$ of (13).  We start with the Einstein contribution
(11a); using the prescription (9), we obtain
\be
\textstyle{\frac{1}{2}} \; 
\pa^2_\mn E_{\a\b} =
\textstyle{\frac{1}{9}} \: X + \textstyle{\frac{1}{9}} \:
Z + T \; .
\ee
In terms of $B$ and $S$ this can be rewritten as
\be
\textstyle{\frac{1}{2}} \; \pa^2_\mn E_{\a\b} =
\frac{2}{9} (B-\frac{1}{4} S),
\ee
the original relation
proposed in \cite{002b}.  The $ \pa^2_\mn L_{\a\b}$ 
of (11b) contains more terms but
is equally straightforward.  We find for it
\be
\pa^2_\mn
L_{\a\b} =
-\textstyle{\frac{1}{9}} \: X + \textstyle{\frac{8}{9}} \:
Z + 2T \; ,
\ee
or equivalently
\be
\pa^2_\mn L_{\a\b} =
+ \textstyle{\frac{1}{9}}(7 B +\textstyle{\frac{1}{2}}S) \; .
\ee
Adding (17) and (19) yields (14); hence the promised formula,
\be
B_{\a\b\mn} = \pa^2_\mn \left(L_{\a\b} + \textstyle{\frac{1}{2}} \:
E_{\a\b} \right)\; .
\ee
In contrast to (18), there is no remainder here.
Incidentally, although (21) has been derived for
D=4, where there is a unique $B$, it can be extended
to any $D$ using the $B$ of (4).

There are two immediate questions about (21):  How unique is our
result, and if it is, why just this combination of $t_{\a\b}$?
We have no wisdom as to the latter: the invariant amplitude provides
no hints about why precisely $(L_{\a\b} + \textstyle{\frac{1}{2}} \:
E_{\a\b})$ requires no remainder in this frame.
To quantify the uniqueness aspect, we 
will write down all superpotentials in RNC, 
since these cover all $t_\ab$.
There are both symmetric and non-symmetric $\D_\ab$
(relevant because we used $E_{\a\b}\neq E_{\b\a}$). 
Symmetric ones are of the form
$S^{\a\b} \equiv \pa^2_{\l\s}$ $H^{\a \l\b \s}$, where
$H$ has the algebraic symmetries of the Riemann tensor;
this ensures both identical conservation and symmetry of
$S^{\a\b}$.  Nonsymmetric ones will be conserved on
only one index, $A^{\a\b} \equiv \pa_\l H^{\a \l\b}$
where $H^{\a \l\b}$ is only antisymmetric in $(\a \l )$.
Since we are concerned in RNC with terms of the form
$\pa^2 g \; \pa^2 g$ and in particular with 
$\pa^2_\mn S^{\a\b}$, $H^{\a \l \b \s}$ depends only
on the metric, while $H^{\a \l\b}$ will be
$\sim g \pa g$.  As shown in the Appendix, 
there just are enough independent superpotentials to
span the complete $(X,Y,Z,T)$
basis and thus to express $B$ uniquely.  Hence, there
is just one effective $t_\ab$ -- in RNC -- that fulfills
our relation, though $t_\ab$'s that are intrinsically different
in an arbitrary frame  may degenerate to a single one in RNC.

\noindent{\bf 4. ~Summary}

Our modest result is that on shell, at the origin of
the RNC frame, there is an exact local equality
between the Bel--Robinson tensor and the double gradient
of a particular energy pseudotensor.  We have conjectured
this relation to be
the ``tensor square root" of a more physical one in tree
level graviton-graviton scattering, whose amplitude
is simultaneously proportional to the square of 
$B_{\mn\a\b}$ and (essentially, if not quite gauge
invariantly) to that of a pseudotensor. 
Despite all this fine print, one cannot
help but wonder if there is more to be
learned from (21).

\noindent {\bf Acknowledgement}

This work was supported by NSF grant PHY93--18511.

\newpage

\renewcommand{\theequation}{A.\arabic{equation}}
\setcounter{equation}{0}

\noindent{\bf Appendix: Superpotentials}

We enumerate here the independent superpotentials in RNC.
Our purpose is to show that they in fact constitute
a basis equivalent
to the $(X,Y,Z,T)$, in terms of which we know that our 
class of tensors
is expandable.  Consider first the symmetric
ones; we are actually concerned with
\be
\D^s_{\mn\a\b} \equiv \pa^2_\mn \, S_{\a\b}
= \pa^4_{\mn \l\s} \: H^{\a \l\b \s}(g) \; .
\ee
As explained in text, $H$ contains no derivatives.
It is then of the form
\be
H^{\a \l\b \s} \equiv (g_{\a\b} g_{\l\s} - g_{\a \s}
g_{\b \l}) K(g)
\ee
where $K$ is a ``scalar" like 1 or $\sqrt{-g}$.
The four derivatives will then all fall on the
$(gg-gg)$ term, or all on $K(g)$
to form the surviving $\pa^2 \: \pa^2 g$
combinations, but not two on
each: a $\pa^2 K(g)$ is excluded
because it would necessarily be proportional
to $g^{\l\s} \pa^2 g_{\l\s}$ which vanishes on shell,
irrespective of the $\pa^2$ indices.  When all
four derivatives fall on the $(gg-gg)$, the
undifferentiated $K$ is a constant, $K (g=\eta_{\mn})$,
and it is easy to see that this gives rise to
the contribution
\be
9\D^s_1 = (-5 X + 2Y +4 Z) K(\eta ) \; .
\ee
When all derivatives act on $K$ alone, we have
the usual transverse projector in $(\a\b )$,
\be
\D^s_2 = \left( \eta_{\a\b} \Box - \pa^2_{\a\b} 
\right) \pa^2_\mn \left( g^{\l\s} g_{\l\s} \right)
\ee
where we have kept the relevant, quadratic in 
metric, part of $K$; two derivatives 
are to be distributed
on each metric, neglecting terms such as
$\Box g^{\l\s} \pa^2_\mn g_{\l\s}$ that vanish on
shell.  The significant point for us is the
presence of a (unique) trace term here, {\it i.e.},
that $\D^s_2$ includes a part
\be
\eta_{\a\b} g^{\s\t,\l\m} g_{\s\t,\l\n} \sim T \; ,
\ee
using (9), the identity (6) and the definition
(13b).  As we shall next see, there are enough
independent antisymmetric contributions to
span the remaining (non-trace) 
basis members $(X,Y,Z)$
between them.  The three possible forms are given by
\be
A^{1}_{\a  \l \b} = g_{\l\s} g_{\a\b ,\s} - g_{\a \s}g_{\l\b ,\s}
\;\;\;\;\;
A^{2}_{\a \l\b} = (g_{\l\s,\a}  - g_{\a \s,\l})g_{\b \s}
\;\;\;\;\;
A^{3}_{\a \l\b} = g_{\a \s} g_{\b \s ,\l} - g_{\l\s} g_{\b \s ,\a}\;.
\ee
In principle there can again be coefficients
$K(g)$ here, but in fact they will not contribute:
We are interested in expressions $\sim \pa^3_{\mn \a}
(g\pa g \: K)$.  Only the $\pa^3
(g\pa g ) \: K(\eta )$ part fails to vanish: 
But as in the symmetric case, $\pa^2K(g) \sim
\pa^2 (g^{\l\s}g_{\l\s})$ always vanishes on shell.
Hence we have the three possible terms
$\D^i_A \equiv \pa^2_\mn A^i_{\a\b}$ with the $A^i$
of (A.6).  The result is as follows
\be 
9\D_A^{1} = -5 X
+ 2Y + 4Z = 9\D_1^s \; ,
\;\;\;\;\;
9\D_A^{2}  =  3 X - 6Z\; ,
\;\;\;\;\;
9\D_A^{3}  = -4X + Y
+ 5Z
\; .
\ee
Being linearly independent, these $\D^i$ are equivalent
to $(X,Y,Z)$; they contain no trace, which is instead
carried by $\D^s_2$, as explained above.  Thus, in RNC
the set of four possible superpotentials, like the
$(XYZT)$, span all (double gradients of) 
the pseudotensors $t_\ab$ as well as $B$.

\end{document}